\documentclass[10pt,final,journal,twocolumn]{IEEEtran}
\usepackage{etex}
\usepackage{cite,graphicx,amsmath,amssymb}
\usepackage{subfigure}
\usepackage{times}
\usepackage{latexsym}
\usepackage{graphicx}
\usepackage{bm}
\usepackage{caption}
\usepackage{stfloats}
\usepackage{cases}
\usepackage{pgf}
\let\labelindent\relax
\usepackage{enumitem}
\usepackage{setspace}
\usepackage{fancyhdr}
\usepackage{amscd}
\usepackage{pst-all}
\usepackage{bm}
\usepackage{booktabs} 
\usepackage{algorithmic}
\usepackage{algorithm}
\usepackage{multirow}
\usepackage{flafter}

\usepackage{fullpage}
\newcommand{\subparagraph}{}
\usepackage{titlesec}
\usepackage{booktabs} 
\usepackage{multirow}
\usepackage{array}
\usepackage{bbm}
\usepackage{hyperref}
\usepackage{cleveref}

\allowdisplaybreaks

\newtheorem{thm}{Theorem}

\usepackage[T1]{fontenc}

\newcommand{\ith}{i^{\text{th}}}
\IEEEoverridecommandlockouts

\begin{document}
\thispagestyle{empty}
\title{Throughput Analysis and User Barring Design for Uplink NOMA-Enabled Random Access}
\author{Wenjuan Yu, \IEEEmembership{Member, IEEE}, Chuan Heng Foh, \IEEEmembership{Senior Member, IEEE}, Atta ul Quddus,\\ Yuanwei Liu, \IEEEmembership{Senior Member, IEEE}, Rahim Tafazolli, \IEEEmembership{Senior Member, IEEE}\\
\thanks{W. Yu, C. Foh, A. Quddus and R. Tafazolli are with the 5G Innovation Centre, Institute for Communication Systems, University of Surrey, Guildford, GU2 7XH, U.K. (Emails: \{w.yu, c.foh, a.quddus, r.tafazolli\}@surrey.ac.uk).}
\thanks{Y. Liu is with the School of Electronic Engineering and Computer Science at Queen Mary University of London, London, U.K. (Email: yuanwei.liu@qmul.ac.uk)}}\maketitle
\begin{abstract}
Being able to accommodate multiple simultaneous transmissions on a single channel, non-orthogonal multiple access (NOMA) appears as an attractive solution to support massive machine type communication (mMTC) that faces a massive number of devices competing to access the limited number of shared radio resources. In this paper, we first analytically study the throughput performance of NOMA-based random access (RA), namely NOMA-RA. We show that while increasing the number of power levels in NOMA-RA leads to a further gain in maximum throughput, the growth of throughput gain is slower than linear. This is due to the higher-power dominance characteristic in power-domain NOMA known in the literature. We explicitly quantify the throughput gain for the very first time in this paper. With our analytical model, we verify the performance advantage of the proposed NOMA-RA scheme by comparing with the baseline multi-channel slotted ALOHA (MS-ALOHA), with and without capture effect. Despite the higher-power dominance effect, the maximum throughput of NOMA-RA with four power levels achieves over three times that of the MS-ALOHA. However, our analytical results also reveal the sensitivity of load on the throughput of NOMA-RA. To cope with the potential bursty traffic in mMTC scenarios, we propose adaptive load regulation through a practical user barring algorithm. By estimating the current load based on the observable channel feedback, the algorithm adaptively controls user access to maintain the optimal loading of channels to achieve maximum throughput. When the proposed user barring algorithm is applied, simulations demonstrate that the instantaneous throughput of NOMA-RA always remains close to the maximum throughput confirming the effectiveness of our load regulation.
\end{abstract}
\section{Introduction} \label{sec:introduction}
Massive machine type communication (mMTC) has been considered to be an imperative and challenging scenario in future communication networks, where millions of Internet of Things (IoT) devices per square kilometre are deployed to support the massive connectivity. It is predicted that machine-to-machine (M2M) connections will experience a noticeable growth, where the number of M2M connections will reach 3.9 billion by 2022 \cite{Cisco2019}. Sporadic access will be expected by a large quantity of devices to transmit small data payloads, which results in an unknown, random subset of active devices at a given transmission instance\cite{CBockelmann2018,PPopovski2018}. One of the major issues is how to utilize the limited radio resources to support a large number of devices transmitting small amounts of data, while maintaining low-latency access. Hence, 3GPP introduced Long Term Evolution (LTE) Category M1/M2 user equipment types and narrowband IoT (NB-IoT) to provide the massive connectivity for wide-area converage, while non-3GPP technologies such as LoRa and Sigfox are also in operation as competing technologies\cite{SMoon2018}.

We shall take the NB-IoT as an example. For the uplink transmissions, a contention-based random access (RA) procedure is performed for initial uplink grant which includes the four-step message handshake between IoT devices and evolved node base station (eNB). Firstly, in step (i), an IoT device starts its RA procedure by transmitting a preamble, i.e., Msg1, to the eNB on the Narrowband Physical RA CHannel (NPRACH). A preamble is composed of four symbol groups and each of them is transmitted on a different subcarrier determined by the fixed size frequency hopping \cite{3GPP211}. If two or more IoT devices randomly choose the same initial subcarrier, the preamble sequence will collide but the eNB can still detect the preamble\cite{LFeltrin2019,NJiang2018}. Then,  in step (ii), the devices that transmit the same preamble successfully detected by the eNB will receive the same RA response (RAR) message, i.e., Msg2, which contains the uplink resource grant and synchronization information. This causes the message transmission collision in step (iii), because those devices that selected the same initial subcarrier in step (i) will transmit the radio resource control (RRC) connection requests, i.e., Msg3, on the same Narrowband Physical Uplink Shared CHannel (NPUSCH) resources\cite{NJiang2019}. Then, the eNB will fail to decode multiple RRC connection requests which results in backoff in the time domain.

In the mMTC scenario, preamble collision will be expected to occur more frequently compared to the traditional IoT scenarios due to the fact that a huge volume of low-cost, low-energy-consumption devices aim to access the limited amount of radio resources simultaneously. According to the RA procedure in NB-IoT network, it can be noticed that the increase of preamble collision may result in severe network congestion and long access delay. On the other hand, in recent years, non-orthogonal multiple access (NOMA) has shown to be a promising multiple access technique for future communication networks, due to its high spectral efficiency by allowing multiple users to simultaneously occupy the same radio resource, such as subcarrier/channel, code and time slot\cite{YLiu20171,ZDing2017,YLiu2016}. For power-domain NOMA, superposition coding is conducted at the transmitter side to perform user-multiplexing, while multiuser separation techniques such as successive interference cancellation (SIC) can be applied at the receiver side to decode the superimposed signals\cite{WYu2019}. Hence, to resolve the severe network congestion and to reduce access latency in mMTC scenarios, in this article we adopt NOMA\footnote{Note that the existing dominant NOMA schemes fall into two categories: power-domain NOMA and code-domain NOMA. In this article, we focus on the power-domain NOMA, which is hereinafter referred to as NOMA.} as an enabling technology which can enhance the conventional ALOHA-based RA procedure in NB-IoT and LTE-M, by allowing multiple IoT devices to transmit messages using the same uplink resource\cite{SMRIslam2017,YLiang2017,SZhou2018}.  

The NOMA-based random access, namely NOMA-RA, allows the users to randomly select one channel and one power level to transmit, by defining a set of pre-determined power levels\footnote{Here, the power level is the received-power level and we adopt the same power model given in \cite{JChoi20171}.}. Then, the eNB with SIC technique will successively decode the received information, based on the received-power difference. Compared to the conventional RA schemes such as pure ALOHA and slotted ALOHA, this newly proposed NOMA-RA scheme is more spectrum-efficient. In \cite{JChoi20171}, by applying the concept of NOMA to multi-channel slotted ALOHA (MS-ALOHA), the author proposed a lower bound for throughput defined as the average number of signals that can be successfully decoded. Considering a NOMA-RA scheme, an approximate throughput was given in \cite{JChoi2018} which takes into account packet collision as well as decoding errors due to low signal-to-interference-plus-noise ratio (SINR). The authors in \cite{JChoi2018} also considered the contention resolution repetition diversity for the NOMA-RA scheme, which guarantees reliable transmissions. However, for the NOMA-RA scheme, the theoretical analysis and the exact expression for throughput and optimal load that gives maximum throughput are not available in the literature, although a lower bound for the throughput is provided in \cite{JChoi20171} which is accurate only when the number of power levels is two. Furthermore, the performance gain of NOMA-RA in terms of throughput is not analyzed, in comparison with the conventional RA schemes with and without capture effect. With capture effect, the receiver has the ability to decode the strongest signal even if there are packet collisions on a given time-frequency resource\cite{ROLaMaire1996}. 

In this article, we focus on a novel NOMA-RA scheme and theoretically study its performance gain in terms of throughput and optimal offered load for the very first time. It is well known that the maximum throughput of conventional slotted-ALOHA is approximately 0.368 achieved when the offered load\footnote{The offered load is defined as the expected number of packets attempted in a time slot.} is 1, which indicates that successful transmissions happen only 36.8\% of time in slotted-ALOHA\cite{ASTanenbaum2010}. Correspondingly, one interesting question that needs to be addressed is: what are the maximum throughput and optimal load for the NOMA-RA scheme? This paper addresses this problem and provides exact analytical expressions for throughput and optimal load that achieves maximum throughput, as well as the comparison results with the conventional MS-ALOHA (with and without capture effect). Our analysis shows that the NOMA-RA scheme provides quite considerable performance gain, even compared to the MS-ALOHA with capture effect.

Apart from designing a more spectrum-efficient RA scheme, another challenging issue for mMTC lies in the traffic burstiness, which is critical when a large number of devices simultaneously activate. To deal with this problem, many techniques have been developed, such as backoff-based mechanisms\cite{DGJeong1995}, access class barring (ACB)\cite{3GPP22011}, and preamble barring\cite{MGrau2019}. Backoff mechanisms defer the re-transmissions of collided packets by a random time to prevent successive collisions. However, as the number of re-transmission attempts increases, the backoff delay may increase exponentially. According to ACB proposed by 3GPP\cite{3GPP22011}, the eNB periodically broadcasts barring parameters including an access probability and a barring duration. Then, all the devices defer their random access requests with the access probability for a barring duration\cite{3GPP22011}. It can be noted that in ACB, it is crucial to control the access probability, which is also an open issue for the newly proposed NOMA-RA scheme. In this article, we focus on the user barring technique which distributes the access attempts according to an access probability for a fixed barring duration. The main contributions of this article are summarized below. 
\begin{itemize}
    \item Focusing on the NOMA-RA scheme, we derive the exact analytical expressions for throughput by considering two different arrival models, i.e., Binomial and Poisson arrivals. Our throughput analysis reveals that an increase in the number of power levels leads to an increase in maximum throughput, but the rate of increase is slower than linear, due to the "higher-power dominance" effect in power-domain NOMA. Further, the optimal load which achieves the maximum throughput is also derived for Poisson arrivals.
    \item A comprehensive comparison between NOMA-RA and the conventional MS-ALOHA (with and without capture effect) is conducted, by utilizing the exact analytical expressions. The performance gain, defined as the ratio of maximum throughput achieved by NOMA-RA to that of MS-ALOHA\footnote{Note that by default, MS-ALOHA refers to the scheme without the consideration of capture effect.}, is illustrated. It is shown that when there are four power levels, the maximum throughput achieved by NOMA-RA triples that of MS-ALOHA.
    \item To alleviate the traffic burstiness, a practical user barring algorithm is proposed for NOMA-RA where two steps are involved: (i) load estimation; (ii) adaptive user access control. The load estimation requires the observation of channel outcomes. We find that traffic load can be accurately estimated based on the instantaneous throughput and idle channels\footnote{We define an idle channel as one that at least one idle power level is observed during the SIC decoding.}. Precisely, the eNB observes the instantaneous throughput and the number of idle channels for a fixed barring period, which are then compared with the analytical results to estimate the current load. Based on the load estimate, the access probability can be adjusted for the next period to achieve optimal load. Simulation results indicate that with the user barring algorithm applied, the instantaneous throughput always remains close to the maximum throughput. 
 \end{itemize}

The remainder of this paper is organised as follows. The system model is first introduced in Section \ref{sec:model}, which discusses the proposed NOMA-RA scheme, as well as the conventional MS-ALOHA. In Section \ref{sec:analysis}, we conduct the theoretical analysis on throughput and optimal offered load for the NOMA-RA scheme, which paves the way for the user barring access design. Based on the analytical results, the adaptive access control is studied and the pseudocode for a proposed user barring algorithm is provided in Section \ref{sec:barring}. Simulation results are included in Section \ref{sec:simulations}, followed by conclusions summarized in Section \ref{sec:conclusions}.

\section{System Model}\label{sec:model}

Let us consider the uplink transmissions for a system consisting of one eNB, $N$ orthogonal channels\footnote{Here, one channel means one frequency resource, e.g., one subcarrier in NB-IoT. Since this paper mainly focuses on the theoretical throughput based on probability analysis, the system model is more general, which is not limited to NB-IoT standards.} and $U$ users. Time is discretized into slots of duration $\tau$ corresponding to the packet transmission time. All devices are assumed to be synchronized to the slot boundary. At the beginning of each transmission slot, all users have the same access probability $P_{\text{access}}$ to transmit a single packet, where $P_{\text{access}}\in[0,1]$. Then, within each slot, the transmission attempts can be modelled as $U$ independent Bernoulli trials. The number of transmitted packets (or the number of active users) in one time slot, $U_{\text{access}}$, follows a Binomial distribution with the expectation $E[U_{\text{access}}]=UP_{\text{access}}$. For NOMA-RA scheme, each active user randomly selects one channel and one power level to transmit. There are $L$ power levels in total, denoted by $\gamma_1 > \gamma_2\dots >\gamma_L > 0$, in an ordered manner. Assume that user $k$ chooses the power level $l$ and the channel $n$ for random access. Then, according to \cite{JChoi20171}, the user $k$'s transmission power is given as $\rho_{n,k} = \gamma_l/|h_{n,k}|^2$ where $\gamma_l$ equals to $\Gamma\left(\Gamma+1\right)^{L-l}$ and $\Gamma$ is the target SINR when there exists only one packet at each power level. To apply this power model, it is assumed that the channel gain between user $k$ and eNB, i.e., $h_{n,k}$, is perfectly known at user $k$ as a priori knowledge, so that the transmission power is adjusted to guarantee that the received power level at the eNB is $\gamma_l$. Note that other power control algorithms may also be applicable but are beyond the scope of this paper\footnote{Since the main focus of this paper is not on signal processing, an unsatisfied power level due to the limitation of hardware ability is out of scope of this work. There are some power control algorithms designed in the literature like \cite{JChoi20171} which manage to reduce the transmission power range, but these are beyond the scope of this paper.}. 

Denote by $I_n$ the index set of active users transmitting through the channel $n$. Then, the received signal at the eNB over channel $n$ can be written as
\begin{align}
    y_n = \sum\limits_{k\in I_n} h_{n,k} \sqrt{\rho_{n,k}} s_{n,k} +n_{n},
\end{align}
where $h_{n,k}$, $\rho_{n,k}$, and $s_{n,k}$ represents the channel coefficient, transmit power and signal from user $k$ through the channel $n$, respectively. Further, $n_{n}\sim \mathcal{CN}(0,N_0)$ is the additive white Gaussian noise where $N_0$ is the noise spectral density. Due to the properties of NOMA, one can notice that the NOMA-RA scheme is more spectrum-efficient compared to the commonly utilized RA schemes because it allows multiple users to simultaneously transmit using the same time-frequency resource. As long as the packets occupying the same radio resource arrive at the eNB with different selected received-power levels, the eNB employed with SIC can successively decode all the received signals\footnote{Perfect SIC is assumed, which means that the decoded signal can be perfectly removed without leaving residual interference.}. Note that the conventional RA schemes, such as pure ALOHA and slotted ALOHA, can only support at most one packet on one radio resource, which result in large collision probability and low throughput in heavy load scenarios. The NOMA-RA scheme can address this problem due to the introduction of a new dimension of power-domain. For a NOMA-RA system with $L$ power levels in total, it can support at most $L$ packets on one radio resource, which can effectively reduce the collision probability and result in higher throughput. However, since there is no central management and all active devices just randomly select one channel and one power level, there is still a probability of collision. Hence, it is important to obtain the exact analytical expressions for the achievable throughput for the NOMA-RA scheme. 

Let us focus on the NOMA-RA scheme and take the user $k$ as an example. Assume that for a given time slot, it chooses the channel $n$ and the power level $l$ to transmit. The successful packet transmission of user $k$ occurs (or known as the case of "successful packet") when both the following conditions are satisfied:
\begin{itemize}
    \item Only user $k$'s packet is transmitted on channel $n$ choosing power level $l$.
    \item On channel $n$, all the signals which choose higher received-power levels, i.e., from level 1 to level $l-1$, can be successfully decoded.
\end{itemize}
If more than two packets transmitted on the same channel arrive at the eNB with the same selected power level, this is called "power collision". All the packet decoding on this power level will fail which in turn fails the decoding on all the lower power levels. This is called the issue of "higher-power dominance", due to the fact that SIC is adopted at the receiver. Once the decoding for one power level is successful, the decoded signal is removed before decoding the next level\cite{WYu2018}. But if the decoding on one power level fails, the signals on the lower power levels cannot be decoded due to high interference power. Hence, in order to successfully decode user $k$'s signal, all the signals on higher power levels need to be successfully decoded and removed, then the user $k$'s packet can be decoded by treating the remaining signals on lower power levels as interference. In this work, we consider that the received-power levels are pre-determined in an appropriate way which rules out the possibility of packets on lower power levels destroying the higher power level's decoding \cite{JChoi20171}. Hence, by assuming that user $k$ transmits on the channel $n$ choosing the power level $l$, only the following situations result in the user $k$'s transmission failure\footnote{Note that in this paper, collisions are assumed to be the only source of transmission failure. The decoding errors due to channel outage will be taken into account in our future research.}: 1) On channel $n$, there is more than one packet choosing the power level $l$, where power collision happens; 2) Only the user $k$ transmits on channel $n$ using power level $l$, but some packets transmitted on this channel with higher power levels (from level 1 to level $l-1$) fail the decoding.
\subsection{Conventional MS-ALOHA without Capture Effect}
For comparison purposes, here we also briefly introduce the conventional MS-ALOHA, without capture effect being considered. Focusing on the uplink transmissions for a system with one eNB, $N$ orthogonal channels and $U$ devices. Within each time slot, all $U$ devices have the same access probability $P_{\text{access}}$ to transmit, where the active devices will choose a channel at random\cite{XLi2012}. We take the user $k$ as an example. When capture effect is not considered, the successful packet transmission for user $k$ in a time slot only occurs when it transmits a packet on an idle channel that there is no user transmitting. On the other hand, if two or more packets access the same channel in a given time slot, then there is a collision and the receiver obtains no information. According to \cite{XLi2012}, we note that when $U$ packets are to access $N$ channels, the average number of successfully transmitted packets is given by $U\left(1-\frac{1}{N}\right)^{U-1}$. When there is a large number of users ($U\rightarrow \infty$), Poisson distribution with a parameter $\lambda$ can be used to model the number of packets accessing each channel. Then, the throughput achieved for Poisson arrivals accessing $N$ orthogonal channels is given as $N\lambda e^{-\lambda}$. By taking the first derivative of $N\lambda e^{-\lambda}$ and setting it to 0, the optimal load $\lambda^*$ can be obtained which equals to 1. This indicates that the best system performance is achieved when there is on average one packet attempting to access one time slot, for each channel.
\subsection{Conventional MS-ALOHA with Capture Effect}\label{subsec:capture}
To provide a comprehensive study, here we also introduce and analyze the conventional MS-ALOHA with capture effect considered. For the conventional MS-ALOHA, the receiver has no SIC decoder, but may have the ability to decode the strongest signal even if there are packet collisions on one channel in a given time slot. This is referred to as the "capture effect". In this work, we adopt a perfect capture model, which means that if one of the $U$ transmitters chooses a power level that is larger than that chosen by each of the remaining $U-1$ transmitters, then capture effect occurs and a packet is successfully received by the receiver \cite{ROLaMaire1996}. In contrary, if two or more packets choose the highest power level among all packets, then there is no successful decoding. One can notice that even with capture effect considered, the conventional MS-ALOHA scheme still can only support at most one packet per channel within one given time slot.

Assuming Binomial arrival process on each channel, the throughput for MS-ALOHA with capture effect can be derived, which is then given by\cite{HZhou1998}
\begin{align}\label{eq:capture0}
    T_{cap}& =  N \sum\limits_{U_i=1}^{U} \dfrac{ {U_i \choose 1} \sum\limits_{g=0}^{L-2}\left(L-g-1\right)^{U_i-1}}{L^{U_i}}\nonumber \\
    &\times{U\choose U_i}  \left(\frac{1}{N}\right)^{U_i} \left(1-\frac{1}{N}\right)^{U-U_i}.
\end{align}
On the other hand, if Poisson arrival process with a parameter $\lambda$ is assumed for each channel, the throughput for MS-ALOHA with capture effect is given by
\begin{align}\label{eq:capture1}
    T_{cap}  = N \sum\limits_{U_i=1}^{\infty} \dfrac{  {U_i \choose 1} \sum\limits_{g=0}^{L-2}\left(L-g-1\right)^{U_i-1}}{L^{U_i}}\dfrac{\lambda^{U_i}e^{-\lambda}}{U_i!}.
\end{align}
In Section \ref{sec:simulations}, numerical results are provided to compare the throughput performance between the proposed NOMA-RA and the conventional MS-ALOHA (with and without capture effect).
\section{Theoretical Analysis}\label{sec:analysis}
Based on the above discussions, we now focus on the theoretical analysis and mathematical expressions for throughput and optimal load for the proposed NOMA-RA scheme. For simplicity, we consider that the access probability $P_{\text{access}}$ is fixed and equals to 1, which means that all $U$ devices will access the system and the number of active users $U_{\text{access}}=U$. In Section \ref{sec:barring}, we will consider an adjustable access probability and propose a user barring algorithm to further reduce the collision probability and maintain the maximum throughput. 
\subsection{Throughput Analysis}\label{subsec:throughput}
Assume that there are $U$ packets to independently and uniformly access $N$ channels. For each channel, there are $L$ power levels which can be chosen. Within a given time slot, each packet randomly selects one channel and one power level to access. We define throughput as the average number of successfully transmitted packets that are delivered from all users through all channels, denoted by $T$. Before deriving it, let us first analyze the conditional throughput for one channel, conditioned on the number of packets accessing this channel.  
\begin{thm}\label{thm:1}
Given that there are $U_i$ packets sent to access channel $i$, the conditional throughput, defined as the average number of successfully transmitted packets on this channel, is given by $E[S_i|U_i]=\sum\limits_{S_i=1}^{\min(L,U_i)} S_i P(S_i|U_i)$, where $P(S_i|U_i)$ is the conditional probability of having $S_i$ packets successful on channel $i$, given in \eqref{eq:PSi} at the top of next page.
\begin{figure*}[!t] 
\normalsize
\begin{equation}\label{eq:PSi}
P(S_i|U_i)=\left\{
\begin{array}{ll}
\sum\limits_{g=0}^{L-1-S_i} \dfrac{{U_i \choose S_i} S_i! {S_i+g \choose S_i}} {L^{U_i}}\Big((L-S_i-g)^{U_i-S_i}-(L-S_i-g-1)^{U_i-S_i}\\
-{U_i-S_i \choose 1 } (L-S_i-g-1)^{U_i-S_i-1}\Big)&\multirow{2}*{$\text{if}\ S_i<\min(L,U_i),$}\\
\specialrule{0em}{0.2ex}{0.2ex}
\dfrac{{U_i \choose S_i} S_i! {L \choose S_i}}{L^{U_i}} &\multirow{2}*{$\text{if}\ S_i= U_i\leq L, $}\\
\specialrule{0em}{0.2ex}{0.2ex}
0 &\multirow{2}*{$\text{otherwise.}$}\\
\end{array}\right.
\end{equation}
\hrulefill
\end{figure*}
\begin{IEEEproof}
See Appendix A. 
\end{IEEEproof}
\end{thm}

Here, condition $S_i<\min(L,U_i)$ represents the cases where not all the $U_i$ packets are successful, which implies that there are packet collisions. For these cases, the conditional probability $P(S_i|U_i)$ can be derived by finding the probability that given $U_i$ packets accessing the channel, there are $S_i$ power levels occupied, each with a packet transmission producing $S_i$ successful packets, with some idle\footnote{Here, "idle" means there is no packet transmitted on this power level.} power levels appearing among the $S_i$ successful packets, followed by a power collision occurring at the power level below all the $S_i$ successful packets and those idle power levels if any, and finally all other possible transmissions below the power collision. Our approach to compute this probability involves in further conditioning the power level that the power collision appears. The full derivation of $P(S_i|U_i)$ is given in Appendix A.

For the the condition $S_i= U_i\leq L$, all $U_i$ packets accessing channel $i$ are successful. The conditional probability $P(S_i|U_i)$ can be derived by finding the probability of distributing all $U_i$ packets onto $L$ power levels, where each level only serves at most one packet. 

Theorem \ref{thm:1} focuses on the throughput of a single channel. We now extend Theorem \ref{thm:1} for the study of multiple orthogonal channels. With $U$ packets randomly accessing $N$ orthogonal channels in one time slot, the number of packets that each channel receives follows a Binomial distribution, where the probability that a channel is contended by $U_i$ packets is simply $P(U_i)={U\choose U_i} \left(\frac{1}{N}\right)^{U_i} \left(1-\frac{1}{N}\right)^{U-U_i}$. Using Binomial distribution to describe the number of packets accessing a particular channel, we immediately obtain the following result.
\begin{thm}\label{thm:2}
Consider that there are $U$ packets randomly accessing $N$ channels independently and unbiasedly. The throughput, defined as the average number of packets which are successfully transmitted on all channels in one time slot, is given by
\begin{align}\label{eq:TB1}
    T &= N\sum\limits_{Ui=1}^U E[S_i|U_i] {U\choose U_i}  \left(\frac{1}{N}\right)^{U_i} \left(1-\frac{1}{N}\right)^{U-U_i},
\end{align}
where $E[S_i|U_i]$ is the conditional throughput given in Theorem \ref{thm:1}. 
\end{thm}

The throughput given in Theorem \ref{thm:2} considers a finite number of packets accessing $N$ channels, where the number of packets appearing on a particular channel follows Binomial distribution. However, in mMTC scenarios, a massive number of low-cost, low-energy-consumption devices are expected (that is $U\rightarrow \infty$). With a large number of potential arrivals onto a channel, Poisson arrivals are more appropriate. Hence, we extend our study to Poisson arrivals as follows.
\begin{thm}\label{thm:3}
Assume that the number of packets on each power level follows Poisson distribution with an arrival rate of $\lambda$. Then, the throughput, defined as the average number of packets which are successfully transmitted on all channels in one time slot, is given by
\begin{align}\label{eq:T2}
    T= N \sum\limits_{i=1}^{L} \lambda e^{-\lambda}\left(e^{-\lambda}+\lambda e^{-\lambda}\right)^{i-1}.
\end{align}
\begin{IEEEproof}
See Appendix B. 
\end{IEEEproof}
\end{thm} 

Theorem \ref{thm:2} and Theorem \ref{thm:3} provide the exact analytical expressions for throughput for NOMA-RA scheme, by considering Binomial and Poisson arrivals, respectively. The accuracy of these expressions will be validated in Section \ref{sec:simulations}, as well as the comparison results with the conventional MS-ALOHA. We then focus on finding the optimal load for the NOMA-RA scheme. 
\subsection{Optimal Load Analysis}\label{subsec:optiload}
As discussed earlier, Poisson distribution is more suitable for modelling packet arrivals from a large number of devices such as the mMTC scenarios. Our aim in this section is to find the optimal load which achieves the maximum throughput for NOMA-RA scheme. According to Theorem \ref{thm:3}, we have that $T= N \sum\limits_{i=1}^{L} \lambda e^{-\lambda}\left(e^{-\lambda}+\lambda e^{-\lambda}\right)^{i-1}$. Optimal load is obtained when the first derivative of $T$ with respect to $\lambda$ equals to 0. Hence, by taking the first derivative, we get that 
\begin{align}
    \dfrac{\partial T}{\partial \lambda} &= N  \sum\limits_{i=1}^{L} \dfrac{\partial \lambda e^{-\lambda}\left(e^{-\lambda}+\lambda e^{-\lambda}\right)^{i-1}}{\partial \lambda}\nonumber\\
    & = N  \sum\limits_{i=1}^{L} \dfrac{\partial \lambda e^{-\lambda}}{\partial \lambda} \left(e^{-\lambda}+\lambda e^{-\lambda}\right)^{i-1}\nonumber\\
    &\quad+ \lambda e^{-\lambda} \dfrac{\partial \left(e^{-\lambda}+\lambda e^{-\lambda}\right)^{i-1} }{\partial \lambda}\nonumber\\
    & = N  \sum\limits_{i=1}^{L} \left(e^{-\lambda}-\lambda e^{-\lambda} \right) \left(e^{-\lambda}+\lambda e^{-\lambda}\right)^{i-1} \nonumber\\
    &\quad- (i-1)\left(\lambda e^{-\lambda}\right)^2 \left(e^{-\lambda}+\lambda e^{-\lambda}\right)^{i-2}\nonumber\\
    & = N \sum\limits_{i=1}^{L} \left(e^{-2\lambda}- i\lambda^2 e^{-2\lambda} \right) \left(e^{-\lambda}+\lambda e^{-\lambda}\right)^{i-2}.
\end{align}
By setting $\dfrac{\partial T}{\partial \lambda}$ to zero, the optimal load $\lambda^*$ can be obtained by solving the following equation. \begin{align}
    N \sum\limits_{i=1}^{L} \left(e^{-2\lambda^*}- i(\lambda^*)^2 e^{-2\lambda^*} \right) \left(e^{-\lambda^*}+\lambda^* e^{-\lambda^*}\right)^{i-2} = 0. 
\end{align}
Note that the obtained $\lambda^*$ is the optimal load for one power level, hence for a single channel with $L$ power levels, the optimal load is simply $\lambda^* L$.

The determination of $\lambda^*$ firstly permits us to compute the maximum throughput of NOMA-RA in order to compare with that of the baseline MS-ALOHA. Secondly, knowing how NOMA-RA should be optimally loaded for its best throughput performance, we can design a practical user barring algorithm which adaptively regulates the traffic load to maintain peak throughput performance. In the following section, we shall introduce the proposed user barring algorithm.
\section{User Barring Algorithm}\label{sec:barring}
For mMTC scenarios, traffic overloading on the RA channel is a challenging issue, which can be especially critical when a large number of devices simultaneously activate, e.g., sensors reconnecting after a power outage\cite{MVilgelm2018}. This simultaneous triggering causes bursty arrivals, which can significantly degrade the performance of RA protocols. To alleviate the traffic burstiness, many techniques have been developed, such as the ACB proposed by 3GPP\cite{3GPP22011} and preamble barring \cite{MGrau2019}. In previous sections, the access probability, i.e., $P_{\text{access}}$, was assumed to be fixed and equals to 1 for all users. In this section, we aim to propose a practical user barring algorithm to perform access control by adaptively adjusting the access probability $P_{\text{access}}$ for all devices. 

To be more practical, we assume that the access probability $P_{\text{access}}$ remains fixed within $\Lambda$ time slots and will be updated for the next period of $\Lambda$. Once the access probability $P_{\text{access}}$ is broadcast to all users, each user  decides whether to participate in the channel access for the period immediately after the broadcast with probability $P_{\text{access}}$, or remains silence otherwise. Then, the transmission attempts can be modelled as multiple independent Bernoulli trials, indicating that the number of active users for the next period of $\Lambda$, i.e., $U_{\text{access}}$, follows a Binomial distribution with the expectation $E[U_{\text{access}}]=UP_{\text{access}}$. Note that $P_{\text{access}}$, $U_{\text{access}}$, and $\mathcal{A}_{\text{access}}$ will remain fixed for the next period of $\Lambda$, where $\mathcal{A}_{\text{access}}$ 
is the index set of active users that decided to participate in channel access, i.e., $|\mathcal{A}_{\text{access}}|=U_{\text{access}}$. Then, within each time slot for the period of $\Lambda$, all $U_{\text{access}}$ active users will randomly access $N$ channels and $L$ power levels, following the NOMA-RA scheme. Note that without central management and coordination, the eNB has no prior knowledge of the number of users accessing the system, i.e., $U_{\text{access}}$. It can only guess the current load after receiving and decoding all signals on the uplink RA channel. Hence, the current load needs to be estimated first at the eNB, followed by the adaptive access control for the next period, with the aim of approaching optimal load. 
\subsection{Load Estimation}\label{subsec:loadestimate}
One of the key designs of user barring algorithm is load estimation. Being able to accurately estimate the load permits opportunity to regulate the loading of channel. In our analytical study, we have derived the optimal load for NOMA-RA to achieve peak throughput performance. According to the throughput results of NOMA-RA (see also Fig. \ref{Figure1}), it can be noticed that observing instantaneous throughput alone is insufficient to estimate the load as a lightly and heavily loaded channel can produce the same throughput level causing ambiguity. Fortunately, we find that the appearance of idle power levels can be used to indicate light and heavy traffic conditions, and thus resolving the ambiguity. Besides, the observation of instantaneous throughput and the appearance of idle power levels does not incur additional signaling.

In our design, when the eNB receives and decodes all signals, it counts the number of successful packets over all channels and all power levels. After a period of $\Lambda$, the normalized instantaneous throughput $T_{\text{insta}}$ can be calculated by  $\sum\limits_{t=1}^{\Lambda}\sum\limits_{i=1}^{N}S^{i,t}_{\text{insta}}/N/\Lambda$, where $S^{i,t}_{\text{insta}}$ is the number of successful packets on channel $i$ in time slot $t$. According to Section \ref{subsec:throughput}, we have obtained the analytical expressions for throughput, which provides the average number of packets that can be successfully transmitted. By comparing $T_{\text{insta}}$ with the pre-calculated throughput matrix $\mathcal{T}$, which stores the throughput values for various loading scenarios, we can find two possible estimates for the number of active users, denoted by $\tilde{U}^{1}_{\text{access}}$ and $\tilde{U}^{2}_{\text{access}}$. 

One interesting phenomenon that has emerged is that when the offered load becomes higher, i.e., when there are more packets expected on one channel, the number of power levels that remain unoccupied becomes less. Without loss of generality, we take the channel $i$ as an example and define an indicator variable $Z_i$, $i\in\{1,2,\dots,N\}$, given by 
\begin{equation}\label{eq:Zi}
Z_i=\left\{
\begin{array}{ll}
1\quad & \text{if any idle power level is observed during}\\ &\text{decoding on the } \ith \text{ channel,}\\
0\quad & \text{otherwise.}
\end{array}\right.
\end{equation} 
Here, $Z_i=1$, $\forall i\in\{1,2,\dots,N\}$, indicates that there is at least one idle power level observed during the decoding\footnote{Note that the decoding on one channel finishes when all power levels are successfully decoded or power collision happens.} for the $\ith$ channel. If $Z_i=1$, we label the $\ith$ channel as an "idle channel". Note that according to our definition, an "idle channel" does not necessarily mean that all power levels are idle on this channel. Instead, an "idle channel" indicates that at least one idle power level is observed before collision or before successfully decoding all power levels. The reason of defining "idle channel" in such a way is because when there is power collision, the successive decoding will terminate at the collision level. This means that the eNB can only label this channel relying on the information collected before the collision. Hence, we decide to investigate the probability of the $\ith$ channel being idle, i.e., $P(Z_i=1)$, denoted by $P_{\text{idle}}$, which is then utilized to estimate the current load in the system.

Recall that the number of unoccupied power levels becomes less when there are more packets transmitted. This indicates that the probability of the channel being idle, i.e., $P_{\text{idle}}$, is a monotonically decreasing function with the offered load. Hence, due to the monotonic property of $P_{\text{idle}}$, a unique "load threshold" $P^{\tau}_{\text{idle}}$ can be obtained which is defined as the probability of idle channel achieved at the optimal load, i.e., $P^{\tau}_{\text{idle}}=P_{\text{idle}}|_{\lambda=\lambda^*}$. Apparently, the load threshold $P^{\tau}_{\text{idle}}$ can be used to distinguish between light load and heavy load. Then, when the eNB decodes all the received signals during the period $\Lambda$, it will also observe the relative frequency of idle channels, i.e., $P^{\text{insta}}_{\text{idle}}$, given by $\sum\limits_{t=1}^{\Lambda}\sum\limits_{i=1}^{N}\mathbbm{1}\left(\text{idle}\right)/N/\Lambda$, where $\mathbbm{1}(\cdot)$ is an indicator function. After a period of $\Lambda$, the obtained $P^{\text{insta}}_{\text{idle}}$ will be compared with the pre-calculated threshold $P^{\tau}_{\text{idle}}$ to find the load estimate. If the observed $P^{\text{insta}}_{\text{idle}}$ is larger than the threshold $P^{\tau}_{\text{idle}}$, it reveals that the system has more idle channels with a higher probability. In this case, we consider that the current load is light. If the observed $P^{\text{insta}}_{\text{idle}}$ is smaller than $P^{\tau}_{\text{idle}}$, then we consider it to be in heavy load. In the following theorem, we will provide the analytical expressions for the probability of channel being idle, i.e., $P_{\text{idle}}$, which will be used to find the load threshold $P^{\tau}_{\text{idle}}$.
\begin{thm}\label{thm:4}
Assume that the number of packets on each power level follows Poisson distribution with parameter $\lambda$. The probability of the channel being idle is given by
\begin{align}\label{eq:p_idle}
     P_{\text{idle}}  &\!=\! \left(e^{-\lambda}+\lambda e^{-\lambda}\right)^L \!-\! \left(\lambda e^{-\lambda}\right)^L \!+\! \left(1-e^{-\lambda}-\lambda e^{-\lambda}\right) \nonumber\\
     &\times\left(\sum\limits_{i=2}^{L}  \left(e^{-\lambda}+\lambda e^{-\lambda}\right)^{i-1} - \left(\lambda e^{-\lambda}\right)^{i-1}\right).
\end{align}
\begin{IEEEproof}
See Appendix C. 
\end{IEEEproof}
\end{thm}

Theorem \ref{thm:4} can be applied directly to the case of multiple channels if users independently and unbiasedly choose a particular channel to access at a given time. Next, we shall apply Theorem \ref{thm:4} to compute the load threshold. By inserting the optimal load $\lambda^*$ given in Section \ref{subsec:optiload} into \eqref{eq:p_idle}, the load threshold $P^{\tau}_{\text{idle}}$ can be obtained immediately. Then, by comparing the instantaneous information regarding the number of idle channels with the calculated threshold, the current load can be accurately estimated. 

\subsection{User Barring}
Note that in order to conduct adaptive access control for the NOMA-RA system, there are mainly two steps involved: a) the eNB collects required information to calculate $T_{\text{insta}}$ and $P^{\text{insta}}_{\text{idle}}$ for the last period of $\Lambda$, which is then compared with $\mathcal{T}$ and $P^{\tau}_{\text{idle}}$ to estimate the current load; b) based on the estimated load, the eNB adjusts the access probability $P_{\text{access}}$ for the next period of $\Lambda$, with the aim of maintaining the optimal load. In the above analysis, we have introduced how to utilize the collected information at the eNB to perform load estimation. In the following, we will focus on the second step and investigate how to adjust $P_{\text{access}}$ for the next period. 

Recall that the probability of idle channel $P_{\text{idle}}$ monotonically decreases with the offered load and the threshold $P^{\tau}_{\text{idle}}$ is the probability value obtained at the optimal load. Then, if the instantaneously obtained $P^{\text{insta}}_{\text{idle}}$ is larger than $P^{\tau}_{\text{idle}}$, it indicates that compared to the optimal load scenario, the current system has a higher probability to have more idle channels. In this case, we consider it to be in light load and the access probability can be increased for the next period, given by $P_{\text{access}} = \min(1, P_{\text{access}} \frac{U^*}{ \tilde{U}^{1}_{\text{access}}})$. In contrary, if $P^{\text{insta}}_{\text{idle}}$ is smaller than $P^{\tau}_{\text{idle}}$, we consider the system to be in heavy load and the access probability needs to be reduced for the next period, given by $P_{\text{access}} = \min(1, P_{\text{access}} \frac{U^*}{ \tilde{U}^{2}_{\text{access}}})$. The pseudocode
for the complete user barring algorithm is given in Algorithm 1. 
\begin{algorithm}[!ht]
\caption{User Barring Algorithm}
\label{algorithm:barring}
\begin{algorithmic}[1]
\REQUIRE~~\\
$N$, $L$, $U$, $\Lambda$, $P_{\text{access}}$, $\mathcal{A}_{\text{access}}$, $\mathcal{T}$, $P^{\tau}_{\text{idle}}$, $U^*$. \\
   \FOR{$t \gets 1$ to $\Lambda$}             
        \STATE All $U$ users generate random probabilities and compare with $P_{access}$ to form $\mathcal{A}_{\text{access}}$;
        \STATE Each active user in $\mathcal{A}_{\text{access}}$ randomly chooses a channel and a power level to access;
        \FOR{$i \gets 1$ to $N$}
        \STATE Perform SIC at the receiver;
        \IF {there is no power collision}
        \STATE Count the number of successful decoding, i.e., $S^{i,t}_{\text{insta}}$;
        \STATE Label it as an "idle channel" if any idle power level is noticed;
        \ELSE
        \STATE Count the number of successful decoding, i.e., $S^{i,t}_{\text{insta}}$;
        \STATE Label it as an "idle channel" if any idle level is noticed before collision;
        \ENDIF
        \ENDFOR
    \ENDFOR
    \STATE Calculate the normalized instantaneous throughput $T_{\text{insta}}$, i.e., $\sum\limits_{t=1}^{\Lambda}\sum\limits_{i=1}^{N}S^{i,t}_{\text{insta}}/N/\Lambda$;
    \STATE Calculate the relative frequency $P^{\text{insta}}_{\text{idle}}$, i.e., $\sum\limits_{t=1}^{\Lambda}\sum\limits_{i=1}^{N}\mathbbm{1}\left(\text{idle}\right)/N/\Lambda$;
    \STATE Compare the actual throughput $T_{\text{insta}}$ with $\mathcal{T}$ to find $\tilde{U}^{1}_{\text{access}}$ and $\tilde{U}^{2}_{\text{access}}$;\\
    \IF {$P^{\text{insta}}_{\text{idle}}\geq P^{\tau}_{\text{idle}}$}
    \STATE $P_{\text{access}} = \min(1, P_{\text{access}} \frac{U^*}{ \tilde{U}^{1}_{\text{access}}})$; \COMMENT{\it{ guess it is in light load}}
    \ELSE 
    \STATE $P_{\text{access}} = \min(1, P_{\text{access}} \frac{U^*}{ \tilde{U}^{2}_{\text{access}}})$; \COMMENT{\it{ guess it is in heavy load}}
    \ENDIF 
\ENSURE $P_{\text{access}}$
\end{algorithmic}
\end{algorithm}
\section{Numerical Results}\label{sec:simulations}
In this section, numerical results are provided to validate the accuracy of the proposed analytical expressions in previous sections and further investigate the proposed NOMA-RA scheme. The performance of the proposed user barring algorithm will also be discussed through simulations. Firstly, theoretical analysis for throughput and optimal load of the proposed NOMA-RA scheme will be validated by comparing with Monte Carlo results. To get Monte Carlo results, we let $U$ packets randomly choose power levels and channels. Then, for one experiment, the number of successfully transmitted packets is calculated, while the throughput can be found after conducting a large number of realizations. The analytical expressions for both  arrival models, i.e., Binomial model and Poisson process, will be confirmed. Furthermore, the performance gain of NOMA-RA is also explicitly shown in this section, compared to the conventional MS-ALOHA (with and without capture effect). In the following, we start from the throughput performance for  NOMA-RA, in comparison with the conventional MS-ALOHA without capture effect.
\subsection{Throughput Performance}\label{subsec:fixed}
\begin{figure}[t]
\centering{\includegraphics[width =\linewidth]
{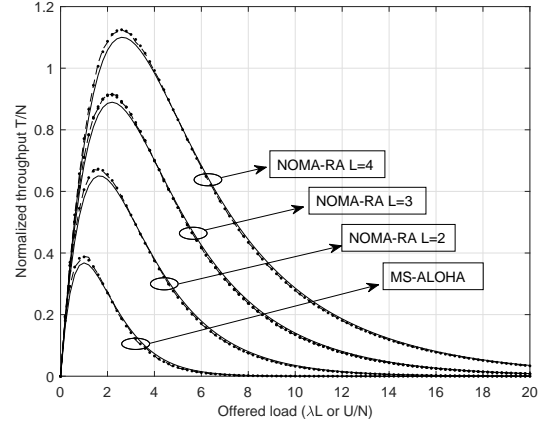}}
\caption{Normalized throughput versus offered load, for NOMA-RA and MS-ALOHA.}\label{Figure1}
\end{figure}
\begin{figure}[t]
\centering{\includegraphics[width =\linewidth]
{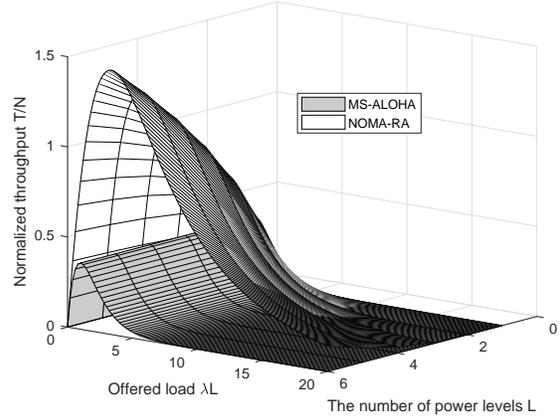}}
\caption{Normalized throughput versus $L$ and $\lambda L$, for NOMA-RA and MS-ALOHA.}\label{Figure2}
\end{figure}

Fig. \ref{Figure1} shows the normalized throughput, defined as $T/N$, versus the offered load ($\lambda L$ or $U/N$), for the proposed NOMA-RA scheme with different number of power levels. Here, the offered load is defined as the average number of packets per channel. For Binomial arrivals, the offered load is $U/N$, while for Poisson arrivals, the offered load per channel is $\lambda L$. The normalized throughput for Poisson arrivals are shown in solid lines, calculated using the analytical expressions given in Theorem \ref{thm:3}. The throughput curves for Binomial arrivals are given using dots, calculated using the expressions given in Theorem \ref{thm:2}, while Monte Carlo results are shown in dashed lines. To plot this figure, it is assumed that $N=10$. Fig. \ref{Figure1} first confirms the accuracy of analytical expressions derived for the throughput for the NOMA-RA scheme, given in Theorem \ref{thm:2} and Theorem \ref{thm:3}. Further, from Fig. \ref{Figure1}, we can note that similar to the conventional MS-ALOHA, the normalized throughput for the proposed NOMA-RA first increases with the offered load and then gradually decreases after reaching its peak value. This confirms that there is one unique maximum throughput and optimal load. When the number of power levels $L=1$, the proposed NOMA-RA scheme is exactly the same with the conventional MS-ALOHA, offering a normalized maximum throughput of 0.368 achieved at the optimal load being 1. When there are more power levels, i.e., $L>1$, NOMA-RA achieves better throughput performance than MS-ALOHA. 
\begin{figure}[t]
\centering{\includegraphics[width =\linewidth]
{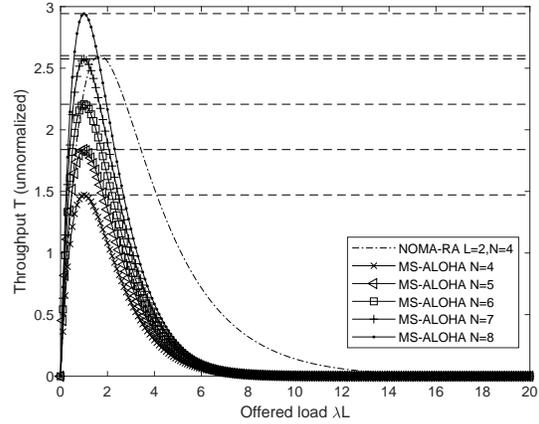}}
\caption{Throughput (unnormalized) versus offered load $\lambda L$, for NOMA-RA and MS-ALOHA.}\label{Figure3}
\end{figure}
\begin{figure}[t]
\centering{\includegraphics[width =\linewidth]
{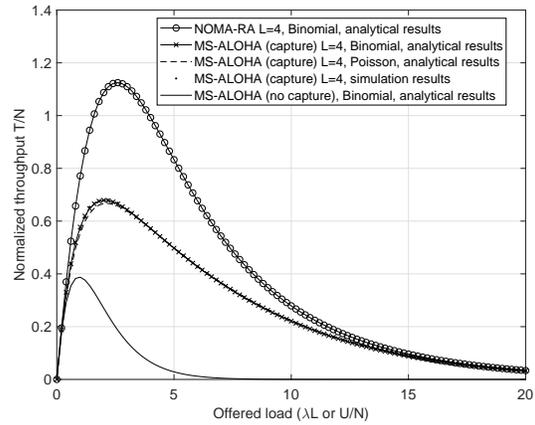}}
\caption{Normalized throughput versus offered load, for NOMA-RA and MS-ALOHA (with and without capture effect).}\label{Figure4}
\end{figure}

To provide more comprehensive studies, Fig. \ref{Figure2} includes the plots for the normalized throughput versus the number of power levels $L$ and the offered load $\lambda$, for the proposed NOMA-RA and the conventional MS-ALOHA. From this figure, we can note that the novel NOMA-RA scheme always achieves higher normalized throughput than the conventional MS-ALOHA, regardless of the given offered load. On the other hand, it further shows that the normalized throughput for NOMA-RA gradually increases with the number of power levels $L$, while that of the conventional MS-ALOHA remains the same. Note that the performance metric considered in this figure is the normalized throughput, defined as $T/N$, which indicates that the impact of multiple channels has been eliminated. This is due to the reason that the linear increase in the number of independent channels, i.e., $N$, results in a linear increase in the throughput (un-normalized). Hence, given a fixed load, the normalized throughput for the novel NOMA-RA scheme gradually increases with the number of power levels, while the normalized throughput remains the same for the conventional MS-ALOHA since its performance increase only relies on the number of channels.

It is noted that the introduction of multiple power levels in NOMA-RA brings another dimension. From the discussions and figures provided above, we can notice that although packet collisions still happen, the proposed NOMA-RA scheme achieves much better throughput performance, compared to the MS-ALOHA (without capture effect). Then, one intriguing question is: how many more channels would be required for MS-ALOHA to achieve the same performance as the novel NOMA-RA? To answer this question, we plot Fig. \ref{Figure3} which includes the curves of throughput (unnormalized), i.e., $T$, versus the offered load, for both schemes. Specifically, it is assumed that there are two pre-determined power levels and four channels available for NOMA-RA. Then, we aim to find the required number of channels for MS-ALOHA to achieve the same throughput performance. Fig. \ref{Figure3} shows that at least eight channels are required for the conventional MS-ALOHA, in order to achieve a similar maximum throughput as the given NOMA-RA scheme. This confirms that the proposed NOMA-RA is much more spectrum-efficient as it only needs half of the radio resources to achieve a similar throughput performance.
\subsection{The Impact of Capture Effect in MS-ALOHA}
The conventional MS-ALOHA discussed above will fail the packet decoding if packet collision happens on one channel, which is a scheme without the consideration of capture effect. According to the perfect capture model discussed in Section \ref{subsec:capture}, if there is one received packet which has a higher power than the others, capture effect occurs and this packet can be successfully decoded. By utilizing the analytical results given in Section \ref{subsec:capture}, it is interesting to check if the novel NOMA-RA is still superior and how much performance gain we can get, compared to the conventional MS-ALOHA with capture effect. Hence, we plot Fig. \ref{Figure4} which aims to compare the normalized throughput for three schemes, i.e., the novel NOMA-RA, the MS-ALOHA with capture effect and the MS-ALOHA without capture effect. From this figure, one can first notice that the proposed NOMA-RA still achieves better performance and the performance gain is quite considerable, even compared to the MS-ALOHA with capture effect. This is mainly because even with capture effect, the conventional MS-ALOHA still can only support at most one packet per channel within one time slot, while the novel NOMA-RA has the possibility to support $L$ packets per channel in one time slot. Hence, the throughput achieved by NOMA-RA can be larger. Fig. \ref{Figure4} also confirms the accuracy of the analytical expressions for the MS-ALOHA with capture effect given in Section \ref{subsec:capture}, which match with the Monte Carlo simulations. 
\begin{figure}[t!]
\centering     
\subfigure[Optimal load]{\label{Figure56:a}\includegraphics[width=\linewidth]{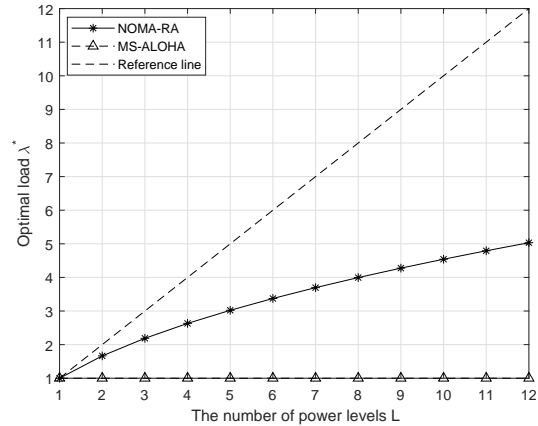}}
\subfigure[Maximum normalized throughput]{\label{Figure56:b}\includegraphics[width=\linewidth]{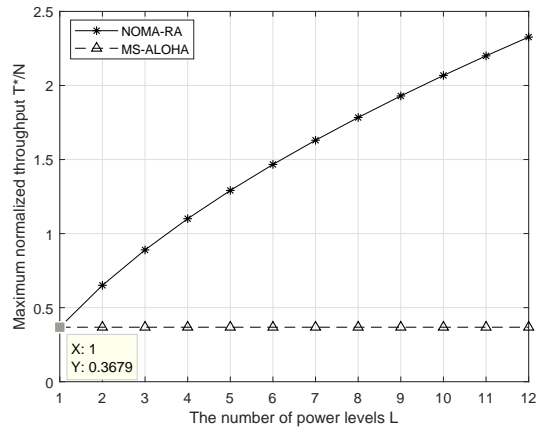}}
\caption{Optimal load and maximum normalized throughput versus $L$.}
\end{figure}
\subsection{NOMA-RA Performance Gain}
In order to investigate the optimal load and the maximum throughput for the proposed NOMA-RA, we include Fig. \ref{Figure56:a} and Fig. \ref{Figure56:b} that are plotted using the analytical results given in Section \ref{subsec:optiload}. The comparison results between NOMA-RA and MS-ALOHA are also given. Fig. \ref{Figure56:a} shows the optimal load $\lambda^*$ versus the number of power levels $L$ for both schemes. 
From Fig. \ref{Figure56:a}, we can first notice that the optimal load for the NOMA-RA scheme starts at the value of 1 and gradually increases with the number of power levels, while for the conventional MS-ALOHA, the optimal load is always achieved at 1. This indicates that the proposed NOMA-RA is more superior than MS-ALOHA since it has the ability to support a larger range of offered load, as we have noticed in Fig. \ref{Figure1}. Interestingly, the optimal load does not increase linearly as the number of power level increases. This is due to the "higher-power dominance" effect in power domain NOMA-RA, resulting from the successive decoding technique. Fig. \ref{Figure56:b} plots the maximum normalized throughput $T^*/N$ versus the number of power levels $L$, for NOMA-RA and MS-ALOHA. As we mentioned above, when $L=1$, NOMA-RA is exactly the same with the conventional MS-ALOHA (without capture effect), which can also be confirmed in Fig. \ref{Figure56:b}. Furthermore, similar to Fig. \ref{Figure56:a}, Fig. \ref{Figure56:b} shows that the maximum normalized throughput for NOMA-RA increases with $L$, while that of MS-ALOHA remains the same. 
\begin{figure}[t]
\centering{\includegraphics[width =\linewidth]
{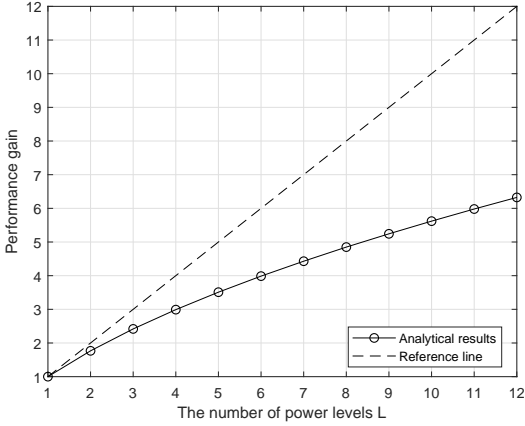}}
\caption{The performance gain of NOMA-RA versus $L$.}\label{Figure7}
\end{figure}

In order to clearly show the performance gain of NOMA-RA, we plot Fig. \ref{Figure7} which shows its performance gain versus the number of power levels. Here, the performance gain is defined as the ratio of maximum throughput achieved by NOMA-RA to maximum throughput supported by MS-ALOHA. Fig. \ref{Figure7} first shows that the performance gain monotonically increases with the number of power levels. To clearly show its increase rate, we provide a reference line plotting the function $y=x$, where $x\in[1,12]$. Fig. \ref{Figure7} shows that the maximum throughput achieved by NOMA-RA triples the conventional MS-ALOHA, when there are four power levels. If we further increase the number of power levels, e.g., $L=6$, the maximum throughput achieved by NOMA-RA is four times larger than that of MS-ALOHA, which can be considered to be a substantial improvement. 
\subsection{User Barring Performance}\label{subsec:adaptive}
Note that we have investigated the proposed NOMA-RA and validated the accuracy of theoretical analysis given in previous sections. In this subsection, we consider that the access probability is adaptive and study the performance of user barring algorithm given in Section \ref{sec:barring}.
\begin{figure}[t]
\centering{\includegraphics[width =\linewidth]
{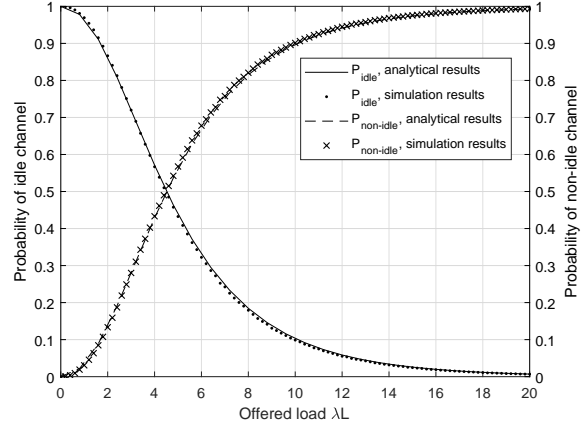}}
\caption{Probability of idle channel and probability of non-idle channel versus offered load.}\label{Figure8}
\end{figure}
\begin{table*}[t!]
\caption{Total number of users in the system} 
\centering 
\begin{tabular}{ c | c |  c | c | c }
  \hline\hline	
  Time & $[0,50\Lambda-1]$ & $[50\Lambda,100\Lambda-1]$ & $[100\Lambda,150\Lambda-1]$ & $[150\Lambda,200\Lambda-1]$  \\[0.5ex]
  \hline
  $U$ & 20 & 50 & 80 & 110 \\
  \hline  
\end{tabular}
\label{table:numberofusers}
\end{table*}
\begin{figure}[t]
\centering{\includegraphics[width =\linewidth]
{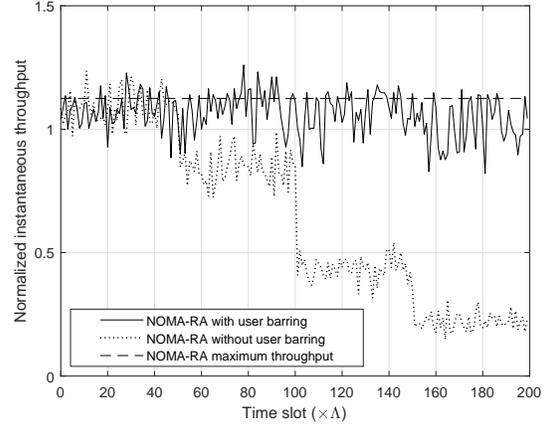}}
\caption{Actual system throughput versus time slot.}\label{Figure9}
\end{figure}
\begin{figure}[t!]
\centering     
\subfigure[Actual load estimate]{\label{Figure1011:a}\includegraphics[width=0.49\linewidth]{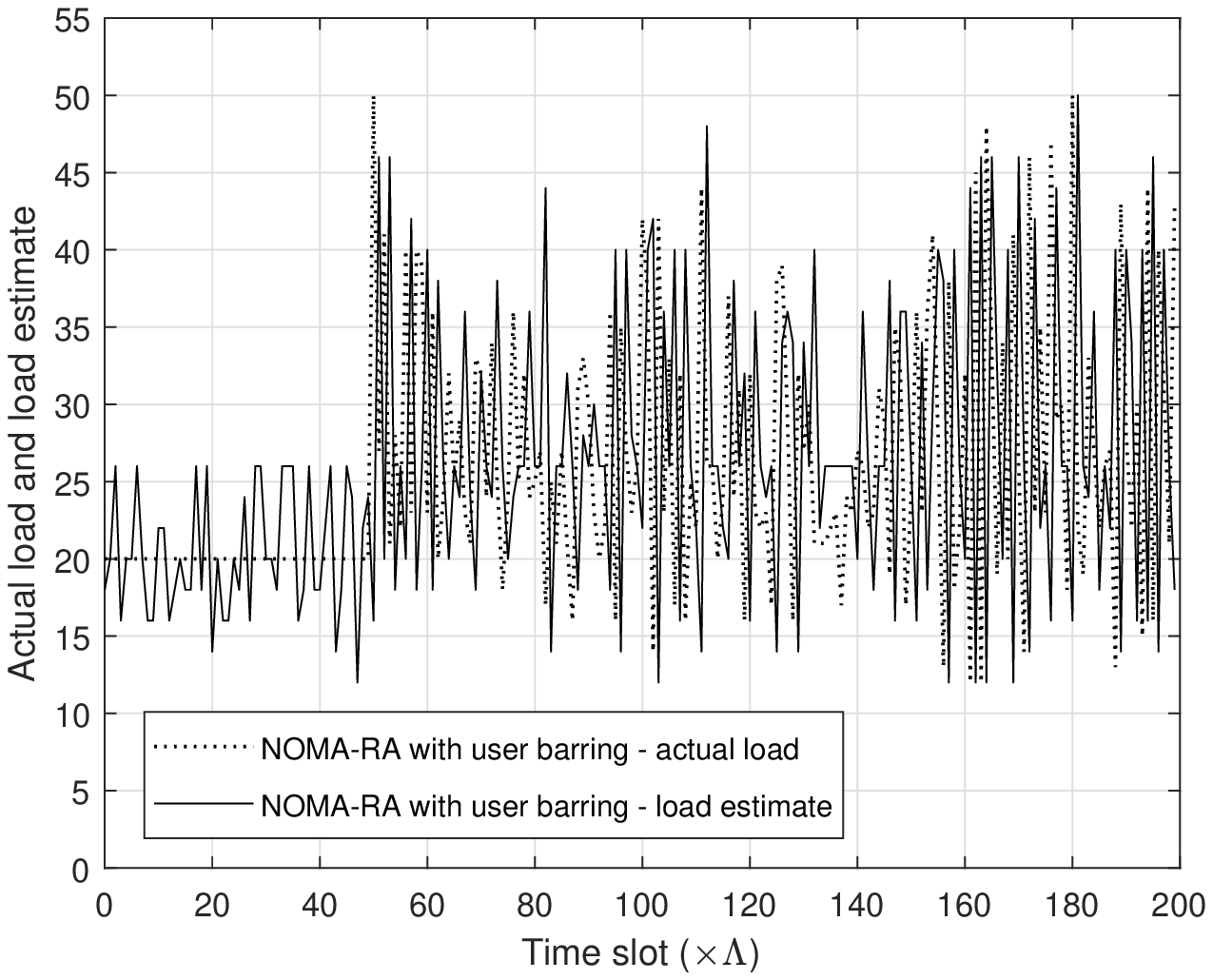}}
\subfigure[Access probability]{\label{Figure1011:b}\includegraphics[width=0.49\linewidth]{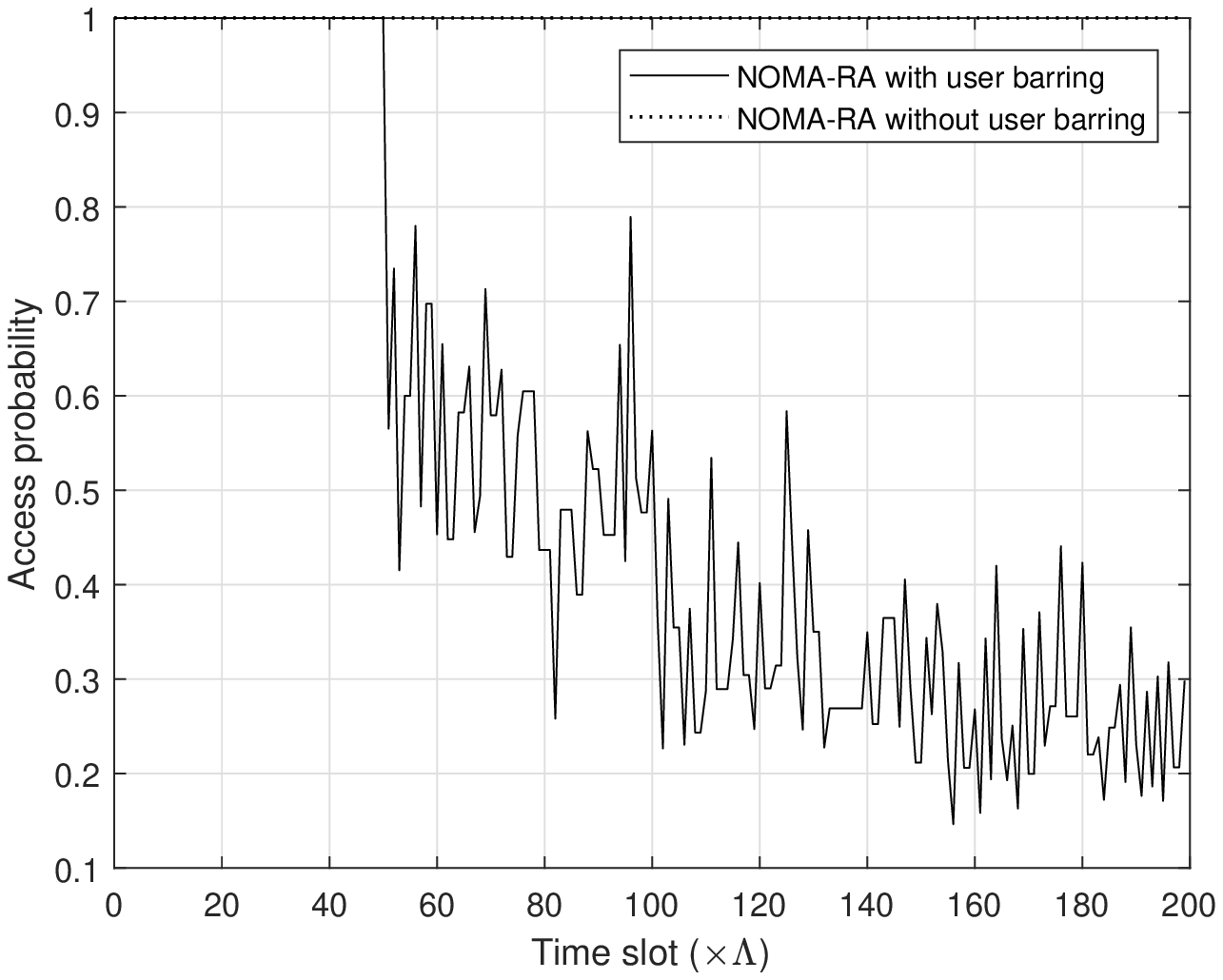}}
\caption{Actual load estimate and access probability versus time slot.}
\end{figure}

Recall that in order to find the accurate estimate for the current load, the probability of idle channel have been introduced. Fig. \ref{Figure8} is plotted to validate the accuracy of analytical expressions for $P_{\text{idle}}$ given in Theorem \ref{thm:4}. To plot this figure, it is assumed that $N=10$ and $L=4$. From this figure, it can be first noticed that the analytical results perfectly match with Monte Carlo simulations, for both probabilities. Further, it shows that the probability of channel being idle starts at the maximum value of 1, gradually decreases, and becomes 0 when the offered load is very heavy. For the probability of non-idle channel, the trend is the opposite. This verifies our initial intention of utilizing the probability of channel being idle/non-idle since it will be a monotonic function with the offered load. Note that the load threshold $P^{\tau}_{\text{idle}}$ utilized in the user barring algorithm is obtained at the optimal load, i.e., $P^{\tau}_{\text{idle}}=P_{\text{idle}}|_{\lambda=\lambda^*}$. According to the analytical results provided in \ref{subsec:optiload}, it can be calculated that when $N=10$ and $L=4$, the optimal load for the NOMA-RA scheme is 2.6. Then, by using the analytical expressions given in Theorem \ref{thm:4}, it can be found that the load threshold $P^{\tau}_{\text{idle}}=0.7822$, 
which can also be confirmed from Fig. \ref{Figure8}.

To study the performance of the proposed user barring algorithm, we provide Fig. \ref{Figure9} which plots the instantaneous throughput for NOMA-RA employed with user barring, in comparison with the actual throughput achieved without user barring and the maximum throughput. To plot this figure, it is assumed that $N=10$ and $L=4$. There are 5000 time slots in total and $\Lambda$ is 25 time slots. According to Algorithm 1, it is noted that $P_{\text{access}}$ and $U_{\text{access}}$ remain fixed during the period of $\Lambda$. Within this period, each active user randomly selects one channel and one power level to access in each time slot. After a period of $\Lambda$, the normalized instantaneous throughput $T_{\text{insta}}$ and the relative frequency of idle channels $P^{\text{insta}}_{\text{idle}}$ are calculated and utilized for load estimate. Hence, it is important to check the performance of $T_{\text{insta}}$ over time. In Fig. \ref{Figure9}, the normalized actual throughput $T_{\text{insta}}$ versus time slot is shown where the total time period is from 0 to $200\Lambda-1$ slots. More specifically, we design a system in which more and more users join every $50\Lambda$ slots as per the details given in Table \ref{table:numberofusers}. From Fig. \ref{Figure9}, it can be noticed that when the offered load changes from light to heavy, the normalized throughput achieved by NOMA-RA with user barring stays close to the maximum throughput, while the throughput achieved without user barring decreases dramatically. This confirms our design intention and proves the effectiveness of the proposed user barring algorithm for NOMA-RA scheme. 

To clearly show the process of load estimate and access probability adjustment, we include Fig. \ref{Figure1011:a} and Fig. \ref{Figure1011:b} which are plotted for the same system and time period with Fig. \ref{Figure9}. Fig. \ref{Figure1011:a} plots the actual load in the system, as well as the estimated load. For the period $[0,50\Lambda-1]$, the number of active users is always 20, equal to the total number of users in the system. This is because initially, $P_{\text{access}}$ is set to be 1 at the time slot 0, which means all 20 users will access and the load is light. The estimated load, i.e., solid line in Fig. \ref{Figure1011:a}, also concludes that it is in light load and therefore $P_{\text{access}}$ remains to be 1 for the period $[0,50\Lambda-1]$. For the period $[50\Lambda,100\Lambda-1]$, the total number of users in the system becomes 50, which is slightly overloaded. If there is no user barring, the access probability $P_{\text{access}}$ will remain to be 1, as we can notice from Fig. \ref{Figure1011:b}. Then, the instantaneous throughput will dramatically decrease since the system is continuously overloaded, which can be confirmed from the dotted line in Fig. \ref{Figure9}. In contrast, with the proposed user barring algorithm applied, the load can be accurately estimated and the access probability $P_{\text{access}}$ is adaptively adjusted which can be observed from the solid lines in Fig. \ref{Figure1011:a} and Fig. \ref{Figure1011:b}. This eventually results in good performance of instantaneous throughput.
\section{Conclusions}\label{sec:conclusions}
This paper focused on the novel NOMA-RA scheme and aimed to study the feasibility and benefits of applying it to mMTC scenario in future communication networks. The advantage of NOMA-RA mainly lies in the introduction of power dimension, which allows multiple devices to transmit with the same time-frequency resource, resulting in higher system throughput. Theoretical analysis and comprehensive comparison studies were conducted, which noticed a substantial improvement in throughput. Furthermore, the NOMA-RA scheme is more spectrum-efficient because it showed that only half of the radio resources is required for a system with two pre-determined power levels to achieve a similar performance, compared to the conventional MS-ALOHA with eight channels. Simulation results showed that NOMA-RA is more superior, even when compared to MS-ALOHA with capture effect. This is because, the conventional MS-ALOHA, with or without capture effect, can only support at most one packet per channel in one time slot, while NOMA-RA can support multiple packets. Finally, with the aim of alleviating traffic burstiness in mMTC, we proposed a user barring algorithm which conducts load estimate and continuously adjusts access probability to perform adaptive access control. Simulation results showed that without user barring, the throughput performance dramatically decreases when the offered load changes from light to heavy, while with the proposed user barring algorithm applied, the instantaneous throughput always remains close to the maximum throughput. 

\section*{Appendix A}\label{app:1}

\section*{Proof for Theorem \ref{thm:1}}
Assume that there are $U_i$ packets to be transmitted on a channel, say, channel $i$. Within a given time slot, each packet randomly chooses one power level to access, where power collisions may happen. In order to calculate the average number of successful packets over $L$ power levels on channel $i$, we first calculate the conditional probability of having $S_i>0$ packets successful, i.e., $P(S_i|U_i)$, where the number of successful packets $S_i$ falls in the range of $[1,\min(U_i,L)]$. We exclude the case $S_i=0$ since this case produces no successful packet and can be omitted when computing the throughput.

To calculate $P(S_i|U_i)$, we consider two cases where the first case includes the presence of power collision and the second case does not contains any power collision. In the first case, we have the condition $S_i<\min(L,U_i)$ indicating that the number of successful packets is less than the number of contending packets and the number of power levels. To derive $P(S_i|U_i)$, we first identify the level where power collision first occurs, then we derive the probability that there are exactly $S_i$ successful packets appearing above the power collision level. If the power collision appears at the $(S_i+g+1)^{\text{th}}$ power level, this indicates that there are $g$ idle power levels appearing above the power collision level. The probability, $P(S_{i,g}|U_i)$, 
that there are exactly $S_i$ successful packets and $g$ idle power levels appearing above the power collision level can be computed by
\begin{align}\label{eq:PSig1}
    &\dfrac{ {U_i \choose S_i} S_i! {S_i+g \choose S_i}} {L^{U_i}}\left((L-S_i-g)^{U_i-S_i}\right.\nonumber\\
    &-(L-S_i-g-1)^{U_i-S_i}- {U_i-S_i\choose 1}\nonumber\\
    &\left.(L-S_i-g-1)^{U_i-S_i-1}\right),
\end{align}
where ${n \choose r}$ is the binomial coefficient. Then, the conditional probability $P(S_i|U_i)$ can be given by taking into account all the possible cases of $g$, i.e., $P(S_i|U_i)=\sum\limits_{g=0}^{L-1-S_i} P(S_{i,g}|U_i)$.

In the second case, we have $S_i = U_i\leq L$ where all $U_i$ packets are successful. The successful probability $P(S_i|U_i)$ can be calculated simply by $\dfrac{ {U_i \choose S_i} S_i! {L \choose S_i}}{L^{U_i}}$. Further, for any other cases satisfying $S_i\in[1,\min(U_i,L)]$ but do not belong to the above scenarios, we have $P(S_i|U_i)=0$. Finally, we obtain the expression for $P(S_i|U_i)$, given in \eqref{eq:PSi}.

We now derive the conditional throughput which is defined as the average number of successfully transmitted packets on channel $i$. Given that there are $U_i$ packets sent to access channel $i$, the conditional throughput $E[S_i|U_i]$ can be determined by $\sum\limits_{S_i=1}^{\min(L,U_i)} S_i P(S_i|U_i)$, by taking into account all possible numbers of successful packets. 

\section*{Appendix B}\label{app:2}

\section*{Proof for Theorem \ref{thm:3}}
Let us focus on a single channel with $L$ power levels. Each power level on one channel can be considered as one virtual resource block, so there are $L$ ordered virtual resource blocks for one channel. Assume that the packet arrival on each power level follows Poisson distribution with a parameter $\lambda$. In other words, the probability of $k$ packets arriving at each power level is given by $q_k = \dfrac{\lambda^k e^{-\lambda}}{k!}$, where $k \in \{0,1,2,\dots\}$. The probability of having a successful packet on the $\ith$ power level (counting from the highest) is simply the probability that a successful packet appears at the $\ith$ power level (which is $q_1$) and no power collision appears at all the above power level (which is $(q_0+q_1)^{i-1}$). Since each power level can carry a successful packet, the number of successful packets a channel can produce is thus $\sum\limits_{i=1}^{L} q_1 (q_0+q_1)^{i-1}$ or $\sum\limits_{i=1}^{L} \lambda e^{-\lambda}\left(e^{-\lambda}+\lambda e^{-\lambda}\right)^{i-1}$.
 When multiple channels are considered, the total throughput $T$ for $N$ channels can be computed by $N\sum\limits_{i=1}^{L} \lambda e^{-\lambda}\left(e^{-\lambda}+\lambda e^{-\lambda}\right)^{i-1}$, due to the fact that all channels are independently and uniformly accessed by users.
\section*{Appendix C}\label{app:3}

\section*{Proof for Theorem \ref{thm:4}}
Let us focus on a particular channel. Given a Poisson arrival with a rate of $\lambda$, the probability of $k$ packets arriving at each power level is given by $q_k = \dfrac{\lambda^k e^{-\lambda}}{k!}$, where $k \in \{0,1,2,\dots\}$. Recall that an "idle channel" is labeled if any idle power level is observed before a power collision or before all power levels are successfully decoded for the case of no power collision\footnote{Please be reminded that a positive number of arrivals are assumed, because when there is no arriving packet, the probability of idle channel is simply one.}. In the following, we shall derive the idle channel probability. 

The probability of having a power collision appearing at a particular level, say, the $\ith$ power level, is $1-q_0-q_1$. To ensure this power collision is the first appearance on the channel counting from the highest power level, all the higher $i-1$ power levels must not contain a power collision. The probability that a power level does not contain a power collision is $q_0+q_1$, and for all the power levels above the $\ith$ to happen, the probability is $(q_0+q_1)^{i-1}$. However, this event includes the case that no idle power level exists above the power collision which should be excluded. This case can happen when each power level above the $\ith$ power level contains a successful packet, with a probability of $(q_1)^{i-1}$. By excluding this case, we have $(q_0+q_1)^{i-1}-(q_1)^{i-1}$. Hence, the probability of the first power collision happening at the $\ith$ power level is given by
\begin{equation}
    (1-q_0-q_1)((q_0+q_1)^{i-1}-(q_1)^{i-1}).
\end{equation}

Finally, considering that the first power collision can occur in any power level except the first which does not allow any idle power level, we get the probability of idle channel, given as
\begin{equation}
    \left(1-q_0-q_1\right)\left(\sum_{i=2}^{L}(q_0+q_1)^{i-1}-(q_1)^{i-1}\right),
\end{equation}
which gives
\begin{equation}
\left(1\!-\!e^{-\lambda}\!-\!\lambda e^{-\lambda}\right)\!\left(\sum\limits_{i=2}^{L}  \left(e^{-\lambda}+\lambda e^{-\lambda}\right)^{i-1} \!-\! \left(\lambda e^{-\lambda}\right)^{i-1}\right)\!.
\end{equation}
The above forms the last term in \eqref{eq:p_idle}.
%

On the other hand, it is possible that there is at least an idle power level but no power collision appearing on the channel. In other words, all power levels must contain either zero or one packet transmission (with a probability of $(q_0+q_1)^L$), and we need to exclude the case where each power level is occupied by exactly one packet transmission which creates no idle power level on the channel (with probability of $(q_1)^L$). With the above, we get the probability of idle channel for this scenario, given by
\begin{equation}
\left(e^{-\lambda}+\lambda e^{-\lambda}\right)^L - \left(\lambda e^{-\lambda}\right)^L,
\end{equation}
which forms the first two terms in \eqref{eq:p_idle}. This completes the proof of \eqref{eq:p_idle} in Theorem \ref{thm:4}.
\bibliographystyle{IEEEtran}
\bibliography{IEEEabrv,bib2014}
\end{document}